\documentstyle[12pt]{ioplppt} 
\jl{6}

\def \a {\alpha}
\def \b {\beta}
\def \se {s_{\rm eff}}
\def \ts {\textstyle}

\begin{document}

\title{Cosmological solutions with nonlinear bulk viscosity}

\author{
Luis P Chimento\dag\,, 
Alejandro S Jakubi\dag\,, 
Vicen\c c M\'endez\ddag\ 
and Roy Maartens\S\
}

\address{\dag\ Departamento de Fisica, Universidad de 
Buenos Aires, 1428~Buenos Aires, Argentina}

\address{\ddag\ Departament de Fisica, Universitat Aut\`onoma 
de Barcelona, Bellaterra~08193, Spain}

\address{\S\ School of Computer Science and Mathematics, Portsmouth
University, Portsmouth~PO1~2EG, Britain}

\begin{abstract}

A recently proposed nonlinear
transport equation is used to model bulk viscous
cosmologies that may be far from equilibrium, as happens
during viscous fluid inflation or during reheating.
The asymptotic stability of the de Sitter and Friedmann solutions is
investigated.
The former is stable for bulk viscosity index $q<1$ and the
latter for $q>1$. New solutions are obtained in the weakly 
nonlinear regime for $q=1$. These solutions are singular and some of 
them represent a late-time inflationary era.

\end{abstract}

\pacs{9880H, 0440N, 0570L, 9880C}

\maketitle

\section{Introduction}

Dissipative processes in isotropic and homogeneous 
Friedmann-Lema\^{i}tre-Robertson-Walker (FLRW) 
universes are restricted to
scalar dissipation by the spacetime symmetries. Such scalar 
dissipation may be analyzed via the relativistic theory of bulk
viscosity (see \cite{m,z} and references cited there).
Relativistic thermodynamics considers
processes which remain close to equilibrium, so that
the transport equation is linear in the bulk viscous stress.
The original and often used 
theory of Eckart,
and a similar theory due to Landau and Lifshitz, are non-causal
(they admit superluminal signals) and their equilibrium
states are unstable \cite{lh}.
The
theory of Israel and Stewart \cite{i,is}, and related 
theories \cite{jcl}, overcome these pathologies, since they 
are causal and 
stable under a wide range of conditions \cite{hl}.

There are dissipative processes in cosmology which are not close
to equilibrium. For example, inflation driven by a viscous
fluid necessarily involves a bulk viscous stress that exceeds
the equilibrium pressure \cite{m,mm}. The reheating era at the
end of inflation also involves far-from-equilibrium dynamics.
In \cite{zpm}, reheating was analyzed via causal thermodynamics.
In that model, the bulk viscous stress is small, and the 
dominant non-equilibrium effects arise from particle creation.
It is also possible to consider models with an increased 
contribution from the bulk viscous stress, beyond the linear regime.

In order to treat these and other
dissipative processes that do not remain close
to equilibrium, one requires a nonlinear generalization of the
standard theories. Recently a nonlinear theory was 
developed in \cite{mm},
via a simple phenomenological generalization of the
Israel-Stewart theory. The nonlinear theory reduces to the causal
theory in the linear regime, and the second law of thermodynamics is 
built into the theory. There is consequently
an upper limit to the bulk viscous
stress, as in nonlinear generalizations of the non-relativistic
heat flow equation \cite{jcl} and shear viscous stress equation
\cite{bjc}.

One can define at each event a local reference equilibrium state with
energy density $\rho$, pressure $p$, number density $n$,
specific entropy $s$ and
temperature $T$. Then the energy-momentum tensor of the fluid 
for the case of scalar dissipation (i.e., no energy or particle
flux and no anisotropic stress) is
\[
T_{\a\b}=(\rho+p+\Pi)u_\a u_\b+(p+\Pi)g_{\a\b}\,,
\]
where $\Pi$ ($\leq0$) is the bulk viscous stress, $u^\a$ is the
four-velocity with respect to which $\rho,p$ and $n$ coincide with
the local
equilibrium values (in FLRW spacetime, this will be the preferred
four-velocity), and $g_{\a\b}$ is the metric (specialized to FLRW
below). The effective non-equilibrium specific entropy is
\cite{jcl,mm}
\begin{equation}
\se=s-\left({\tau\over2nT\zeta}\right)\Pi^2\,,
\label{1}\end{equation}
where $\tau(\rho,n)$ is the characteristic timescale for linear
relaxational effects (the crucial thermodynamic parameter in
Israel-Stewart theory that ensures causality), and $\zeta(\rho,n)$ is
the linear bulk viscosity. 
The evolution of the equilibrium and effective specific entropies 
is given by \cite{mm}
\begin{eqnarray}
\dot{s} &=& -{3H\Pi\over nT}\,, \label{2} \\
\dot{s}_{\rm eff} &=& {\Pi^2\over nT\zeta} \left[1+{\tau_*\over\zeta}
\Pi\right]^{-1} \,, \label{3} 
\end{eqnarray}
where $H={1\over3}\nabla_\a u^\a$ is the Hubble rate 
and $\tau_*$ is a characteristic timescale for nonlinear effects.
The second law of thermodynamics imposes an upper limit on the
bulk stress \cite{mm}:
\begin{equation}\label{repp}
|\Pi| \leq {\zeta\over\tau_*}\,.
\end{equation}
(Note that in the linear theory, there is no upper limit on $|\Pi|$
that is built into the theory -- $|\Pi|/p\ll 1$ is implicitly imposed
a priori since the theory is linear.)

The nonlinear transport equation for $\Pi$ is \cite{mm}
\begin{eqnarray}
&& \tau\dot{\Pi}\left(1+{\tau_*\over\zeta}\Pi\right)+\Pi(1+3\tau_*H)
\nonumber \\
{}&& =-3\zeta H-{\ts{1\over2}}\tau\Pi\left[3H+{\dot{\tau}\over\tau}
-{\dot{\zeta}\over\zeta}-{\dot{T}\over T}\right]\left(1+
{\tau_*\over\zeta}\Pi\right)\,.
\label{4}
\end{eqnarray}
Israel-Stewart theory (in its full, `non-truncated' form \cite{hl,m})
is recovered when $\tau_*=0$. If in addition $\tau=0$, then the
transport equation reduces to that of Eckart.

In a flat FLRW universe, Einstein's equations are\footnote{We
use units with $8\pi G=1=c$ and $k_{\rm B}=1$.}
\begin{eqnarray}
H^2 &=& {\ts{1\over3}}\rho\,, \label{7}\\
\dot{H}+H^2 &=& -{\ts{1\over6}}(\rho+3p+3\Pi) \,,   \label{8}
\end{eqnarray}
so that by (\ref{4}) and (\ref{8}), we get a second-order evolution
equation for $H$ when $\tau\neq 0$.
In order to find this evolution equation for 
nonlinear dissipative processes
in a flat FLRW universe, we need to specify the thermodynamic
parameters $\zeta,\tau$ and $\tau_*$, and the equations of state
relating $p,\rho,T$ and $n$. These will depend on the physical
situation being analyzed, and all quantities except $\tau_*$ follow
from standard thermodynamics. In the nonlinear
theory of \cite{mm}, the microscopic interactions governing
far-from-equilibrium behaviour are unknown and encoded into the 
phenomenological parameter $\tau_*$.
As a first step towards a more complete
analysis, we follow \cite{mm} and
postulate here a simple ansatz for $\tau_*$, which allows
us to investigate the qualitative nature of the effects of 
nonlinearity:
\begin{equation}
\tau_*=k^2\tau\,,
\label{5}\end{equation}
where $k$ is a dimensionless
constant. 
This assumption is a mathematical simplification, but will allow us
to find some overall features of nonlinear viscosity (whose
detailed physics is not known).
The weakly nonlinear regime is then
characterized by $k^2\ll 1$, and we will linearize the evolution
equation for the Hubble rate below. 

The linear relaxation time is related to the bulk viscosity
by the physical relation \cite{m2}
\begin{equation}
\tau={\zeta\over v^2(\rho+p)}\,,
\label{6}\end{equation}
where $v$ is the dissipative contribution to the speed of sound
$V$ ($\leq 1$), 
so that $V^2=c_{\rm s}^2+v^2$, with $c_{\rm s}$ the adiabatic
contribution. 
This relation follows from an analysis of the propagation
of small disturbances in a dissipative fluid.
A simple form for the bulk viscosity is often taken
to be $\zeta\propto \rho^{q/2}$ ($q$ constant), so that by (\ref{7})
we have
\begin{equation}
\zeta=\a H^q \,,
\label{9}\end{equation}
where $\a$ ($\geq 0$) 
is a constant. Although this is a mathematical ansatz
introduced for simplicity, it does approximate the physical form
of $\zeta$ for certain fluids \cite{pbj}.

We also need to specify the equations of state. 
For the pressure, we will take the simple linear barotropic law
\begin{equation}
p=(\gamma-1)\rho \,,
\label{10}\end{equation}
where $\gamma$ is a constant. Typically $\gamma\approx{4\over3}$  
(radiation-dominated expansion) or $\gamma\approx1$ (matter-dominated
expansion).
(Note that the limiting case of dust, when $p$ is exactly 0 and
$\gamma$ is exactly 1, is strictly ruled out, since the temperature
is zero for dust and there can be no bulk viscous stress.)
By (\ref{10}), $c_{\rm s}^2=\gamma-1$ is a
constant, and we will assume that $v$
is also constant, subject to the causality constraint
\begin{equation}
v^2\leq 2-\gamma \,.
\label{13}\end{equation}
This is a mathematical simplification, but it should not be
unreasonable if (\ref{10}) holds.

By equations (\ref{7}) and (\ref{5})--(\ref{10}), we
can use the above thermodynamic assumptions to rewrite the upper
limit (\ref{repp}) on the bulk stress, and to express $\tau_*$
in terms of the Hubble rate:
\begin{eqnarray}
|\Pi|&\leq&{\zeta\over\tau_*}=\gamma\rho\left({v\over k}\right)^2
\leq\left[{\gamma(2-\gamma)\over k^2}\right]\rho\,,
\label{13'}\\
\tau_* &=&\left({\alpha k^2\over3\gamma v}\right)H^{q-2}\,. 
\label{13''}
\end{eqnarray}
It follows from (\ref{13''}) that in the case $q=1$, the nonlinear 
(and linear) timescales are directly determined by the expansion
timescale $H^{-1}$.

Finally, for the temperature,
we will consider two cases: (a) barotropic temperature, $T=T(\rho)$,
so that \cite{m2}
\begin{equation}
T \propto \rho^{(\gamma-1)/\gamma}\,,
\label{11}\end{equation}
and (b) ideal-gas temperature, so that
\begin{equation}
p=nT \,.
\label{12}\end{equation}
Note that (\ref{11}) and (\ref{12}) can only hold simultaneously
in equilibrium \cite{z,m2,mt}.

Equations (\ref{7})--(\ref{10}) together with
(\ref{11}) and the number conservation equation
$\dot{n}+3Hn=0\,$
give the evolution equation for case (a) as \cite{mm}
\begin{eqnarray}\label{14}
&& \left[1-{k^2\over v^2}-\left({2k^2\over3\gamma v^2}\right){\dot{H}
\over H^2}\right]\left\{\ddot{H}+3H\dot{H}+\left(
{1-2\gamma\over\gamma}\right){\dot{H}^2\over H}+{9\over4}
\gamma H^3\right\}  \nonumber\\
&&{}+{3\gamma v^2\over2\alpha}\left[1+\left(
{\alpha k^2\over\gamma v^2}
\right)H^{q-1}\right]H^{2-q}\left(2\dot{H}+3\gamma H^2\right)
-{9\over2}\gamma v^2H^3=0 \,. 
\end{eqnarray}
In case (b), the ideal-gas behaviour
(\ref{12}) leads to the new and simpler evolution equation
\begin{eqnarray}
& &\left(1-\frac{k^2}{v^2}-\frac{2k^2}{3\gamma v^2}
\frac{\dot{H}}{H^2} \right)
\left\{H\ddot{H}-2\dot{H}^2-\frac{9}{2}v^2\gamma H^4 \right\}
\nonumber\\
& &{}+\frac{3v^2\gamma}{2\alpha}H^{3-q}(2\dot{H}+3\gamma H^2)=0 \,.
\label{15}
\end{eqnarray}

Considerable work has been done using the linear non-causal 
model of bulk viscosity (see \cite{pbj,m} for references), 
including the case of cosmic
strings \cite{b1,b2,b3,t}.
Our results represent a generalization to the nonlinear
case of previous dynamical analysis in the linear causal theory
\cite{pbj,mt,chm,Chi96d,luis,cjm}, and also a generalization of
some of the results in \cite{mm}, where
exact de Sitter solutions for any $q$, and 
a power-law solution with $q=1$ were given
for equation (\ref{14}). Here we find new solutions with $q=1$
in the weakly nonlinear regime for both (\ref{14}) and (\ref{15}).
We also investigate asymptotic solutions and their stability
for $q\neq 1$. In principle our solutions and their properties
could be used to model far-from-equilibrium processes, such as
viscous fluid inflation and reheating. 
However, we will not pursue such applications here.

\newpage

\section{Models with barotropic temperature} 

\subsection{Solution for $q=1$}

When the bulk viscosity index satisfies
$q=1$, we have from (\ref{9}) and
(\ref{13''}) that $\zeta\propto H$ and $\tau,\tau_*\propto H^{-1}$.
Assuming that $k^2\ll 1$ and $\dot H/H^2$ is bounded, we can 
isolate $\ddot{H}$ in (\ref{14}) to first order
in $k^2$:
\begin{eqnarray} 
&& H\ddot H+\left(\frac{1-2\gamma}{\gamma}+
\frac{2k^2}{\alpha}\right)\dot H^2
+3\left(1+\frac{\gamma v^2}{\alpha}+\frac{2\gamma k^2}{\alpha}\right)
H^2\dot H \nonumber\\
&&{}+\frac{9\gamma}{2}\left(\frac{1}{2}+\frac{\gamma v^2}{\alpha}
-v^2+\frac{\gamma k^2}{\alpha}\right) H^4=0 \,.
\label{H}
\end{eqnarray}
With the change of variables 
\[
H=y^n\,,~~ t=\theta/(3m) \,,
\]
where
\[
{1\over n}= {(1-\gamma)\over\gamma}+{2k^2\over\alpha} \,,~~
m = 1+\frac{\gamma v^2}{\alpha}+\frac{2\gamma
k^2}{\alpha}\,,
\]
equation (\ref{H}) becomes\footnote{Note that $n=-1$ is ruled out
since $k^2/\alpha$ is non-negative.}
\begin{equation} \label{y}
{d^2y\over d\theta^2}+y^n {dy\over d\theta}
+\frac{\beta}{n+1}y^{1+2n}=0 \,,
\end{equation}
where
\begin{equation} \label{beta}
\beta = \left[\frac{\alpha\left(\alpha+2\gamma v^2-2\alpha
v^2\right)}{4\left(\alpha+\gamma v^2\right)^2}\right]+
\left[\frac{\gamma v^2\left(\alpha^2+\gamma^2 v^2-\alpha\gamma
v^2\right)}{\left(\alpha+\gamma v^2\right)^3}\right]k^2+ O(k^4) \,.
\end{equation}
Then, making the nonlocal transformation \cite{luis}
\begin{equation} \label{ch}
z=\frac{1}{n+1}y^{n+1}\,,~~ d\eta=y^n d\theta \,,
\end{equation}
equation (\ref{y}) becomes linear:
\begin{equation} \label{z}
\frac{d^2z}{d\eta^2}+\frac{dz}{d\eta}+\beta z=0.
\end{equation}
From (\ref{beta}) it is easy to check that $1-4\beta>0$, so we
obtain the general solution of (\ref{H}) in parametric form
\begin{eqnarray}
H(\eta)&=&\left[(n+1)\left\{C_+\exp\left(\lambda_+\eta\right)+
C_-\exp\left(\lambda_-\eta\right)\right\}\right]^{n/(n+1)} \,,
\label{gs0}\\
t(\eta)&=& t_0+\left({1\over3m}\right)\int {d\eta\over H(\eta)}\,,
\label{gs}\end{eqnarray}
where $\lambda_\pm=[-1\pm\left(1-4\beta\right)^{1/2}]/2$ are real, 
and $t_0$,
$C_\pm$ are arbitrary integration constants. When either of
$C_\pm$ vanishes,
we obtain one-parameter families of solutions, which have the form:
\begin{eqnarray}
&&H_{\pm }(t)={\nu _{\pm }\over(t-t_0)}~\mbox{ if }~
 \alpha \neq \alpha_0
\equiv\frac{2\gamma\left(v^2+k^2\right)}{2v^2-1}\,,
\label{1pa}\\
&&H_{-}={\nu _0\over (t-t_0)}\,,~ H_{+}=H_0~\mbox{ if }~
 \alpha = \alpha_0\,,
\label{1pb}
\end{eqnarray}
where 
\begin{eqnarray} 
\nu _{\pm }&=&\left({2\over3\gamma}\right)\frac{m\pm
\left\{m^2-\left[m+\left(\gamma/\alpha-2\right) v^2\right]\left(
1+2\gamma k^2/\alpha\right)\right\}^{1/2}}
{ m+\left(\gamma/\alpha-2\right) v^2} \,, \label{n1}\\
\nu_0 &=&{\a +2\gamma k^2\over 3\a\gamma m}\,, \label{n2}
\end{eqnarray}
and $H_0$ is an arbitrary positive constant giving a de Sitter
solution (with $v>1/\sqrt{2}$), which agrees with the linearization of
the de Sitter solution given in \cite{mm}, equation (41).

The parameter $\nu_-$ is smaller than the perfect fluid value:
$0<\nu_-<2/\left(3\gamma\right)<1$,
and for small dissipative contribution to the
speed of sound we get $\nu_-\approx
2\left(3\gamma\right)^{-1}\left(1-\sqrt{2}v\right)$. Also
$\nu_0<\left(3\gamma\right)^{-1}<1$ for $1<\gamma<2$, 
while $-\infty<\nu_+<\infty$.
Thus, (\ref{1pb}+) contains 
singular solutions without particle horizons.

Evaluating the local-equilibrium entropy $s$ from (\ref{2}), we find
\begin{equation}\label{1se}
s(t) = s_0 + \gamma\left[\frac{
(3\nu_\pm^2)^{1/\gamma}a_0^3}{n_0T_0}\right]
(t-t_0)^{3\nu_{\pm}-2/\gamma}\,,
\end{equation}
for the solution given by (\ref{1pa}), and the same for
(\ref{1pb}--), but
with $\nu_0$ instead of $\nu_{\pm}$.
The effective specific entropy $\se$ is given by (\ref{1}):
\begin{equation}\label{2sef}
\se(t) = s_0 +\left[
\frac{({3\nu_{\pm}^2)}^{1/\gamma}a_0^3}{n_0T_0}\right]\left[\gamma 
-\frac{1}{2v^2\gamma}
\left(\frac{3\gamma \nu_{\pm}-2}{3\nu _{\pm}} \right)^2\right]
(t-t_0)^{3\nu_{\pm} -2/\gamma} \,.
\end{equation}
Hence the requirement $\se\geq 0$ is fulfilled for
\begin{equation}
\frac{3\nu _{\pm}\gamma -2}{3\nu_{\pm}}<\sqrt{2}v\gamma \,.
\label{1r}
\end{equation}
For the solution (\ref{1pb}--), the restriction (\ref{1r}) and the
solution (\ref{2sef}) are the same, but
$\nu_{\pm}$ is replaced by $\nu _0$. Provided that $\nu_0 >0$,
then this restriction requires $v>1/\sqrt{2}$.

The second restriction, arising from positivity of the
entropy production is
\begin{equation}
\frac{3\nu _{\pm}\gamma -2}{3\nu_{\pm}}<\frac{\gamma v^2}{k^2}
\label{2r}
\end{equation}
for the solution (\ref{1pa}), and the same but replacing $\nu_{\pm}$
by $\nu_0$ for (\ref{1pb}). Thermodynamics restrictions (\ref{1r})
and (\ref{2r}) are the same one for $v=\sqrt{2}k^2$.

We are able to find a third solution.
Provided the parameters satisfy the constraint
\begin{equation} \label{mag0}
{\gamma\left[ m+\left(\gamma/\alpha-2\right)v^2\right]\over 8\a m^2}=
{\a^2(1-\gamma)+2\gamma\a k^2\over\left[\a(2-\gamma)
+4\gamma k^2\right]^2}\,,
\end{equation}
the general solution of 
(\ref{H}) takes an explicit form, and we get for the
scale factor
\begin{equation} \label{at}
a(t)=a_0\left| \left|t-t_0\right|^{n+1}+K\right|^{(n+2)/3nm}\,,
\end{equation}
where $K$ is an integration constant.
We have checked that the constraint 
(\ref{mag0}) is satisfied for suitable
values of the parameters.

As examples, we apply 
the two thermodynamical restrictions (\ref{1r}) and (\ref{2r})
and the causality constraint (\ref{13})
to the solutions (\ref{1pa}) and (\ref{1pb}),
for a radiation-dominated expansion, $\gamma={4\over3}$. Causality
demands in this case $v^2 \leq {2 \over 3} $.\\
$\bullet$ For the solution (\ref{1pa}) we choose 
$v^2={1\over2}$ and $\alpha =1$. Causality and the 
two thermodynamical restrictions are
fulfilled for any $k^2 \ll 1$. By (\ref{n1}), it follows that for
small $k$ ($k<0.32$), we have $\nu_+>1$, so that these solutions are always
inflationary, and $\nu_-<1$, corresponding always to non-inflationary
expansion. The maximum value of $\nu_+$ is ${1\over8}(5+\sqrt{13})$,
and the minimum value of $\nu_-$ is ${1\over8}(5-\sqrt{13})$.
Thus $a\sim t^N$ where $0.17<N<1.08$. For small $\alpha$ (perfect
fluid limit) we get $\nu_+ \approx 2(3\gamma)^{-1}<1$ and
$\nu_-\approx (2/3\gamma)(k^2/(k^2+v^2))\ll 1$. Thus the solution
with $\nu_+$ corresponds to the perfect fluid limit solution, while
the other corresponds to a new solution that does not exist in this
limit. On the other hand, for large $\alpha$ we get
$$\nu_{\pm} \approx \frac{2}{3\gamma}\frac{1}{1 \mp \sqrt{2}v} $$
thus, $\nu_{\pm}$ may be very large for $v^2$ close to $1/2$
(expanding solutions require $v^2<1/2$).\\
$\bullet$ For the solution (\ref{1pb}--), we choose $v^2={2\over3}$,
and by (\ref{n2})
\[
\nu_0={\ts{3\over16}}\left(1-{\ts{5\over4}}k^2\right)+O(k^4)\,,
~~\a_0={\ts{4\over3}}\left(2+3k^2\right)\,.
\]
The solution is not inflationary, and causality and thermodynamic
requirements are always fulfilled.

The requirement that $\dot H/H^2$ remains bounded 
means that not all the
two-parameter families of solutions (\ref{gs}) of 
equation (\ref{H}) are also
solutions of (\ref{14}), but it is a consequence of the thermodynamic
restriction given by positivity of the entropy production. So, from
(\ref{repp}), (\ref{7})--(\ref{6}) 
and (\ref{10}),
we find that $\dot H/H^2$ is bounded:
\[
\frac{\dot{H}}{H^2}\leq \frac{3\gamma}{2k^2}(v^2-k^2) \,.
\]
Solutions that satisfy this 
behave as follows \cite{Chi96d}:\\
The evolution begins at a singularity with a Friedmann leading
behaviour of the form (\ref{1pa}--) for $\alpha\neq \alpha_0$ 
or (\ref{1pb}--) for
$\alpha=\alpha_0$. 
Then the expansion becomes

1. asymptotically Friedmann, like (\ref{1pa}+), for $\alpha<\alpha_0$;

2. asymptotically de Sitter, like (\ref{1pb}+), for $\alpha=\alpha_0$;

3. divergent at finite time, with leading behaviour (\ref{1pa}+), for
$\alpha>\alpha_0$.

\subsection{Asymptotic solutions for $q>1$ and stability}

For $q>1$ it is easy to check that (\ref{14}) admits a solution 
whose leading term is $2/(3\gamma t)$ when $t\rightarrow\infty$.  
To study its stability, we make the
change of variables
\[
H^{-1} = t[u(z)]^{\gamma/(\gamma-1)}\,, 
~t^{q-1} = z\,,
\]
when $\gamma>1$. Then (\ref{14}) takes the form
\begin{eqnarray}
&& u''+\left\{\frac{6\gamma v^2}{\alpha R}
u^{\gamma(q-2)/(\gamma-1)} 
+\left[6\left(1+\frac{k^2}{R}\right)u^{\gamma(1-\gamma)}
+q-\frac{2}{\gamma}\right]\frac{1}{z}\right\}\frac{u'}{q-1}\nonumber\\
&&{}+\frac{3\left(\gamma-1\right)v^2}{\left(q-1\right)^2\alpha R}
\left[u^{(\gamma q-\gamma-1)/(\gamma-1)}-\frac{3\gamma}{2}
u^{(\gamma q-2\gamma-1)/(\gamma-1)}\right]\frac{1}{z} \nonumber\\
&&{}\frac{\gamma-1}{\left(q-1\right)^2}\left[-\frac{9}{4}
\frac{2 k^2-2 v^2+R}{R}u^{(\gamma+1)/(1-\gamma)}+
3\frac{k^2+R}{\gamma R}u^{1/(1-\gamma)}-\frac{u}{\gamma^2}
\right]\frac{1}{z^2} \nonumber\\
&&{}=0 \,,
\label{dusg}
\end{eqnarray}
where
a prime denotes $d/dz$ and
\begin{equation} \label{Rsg}
R=1-\frac{k^2}{v^2}+\left(\frac{2k^2}{3\gamma v^2}\right)
u^{\gamma/(\gamma-1)}
\left[1+\frac{\gamma\left(q-1\right)}{\gamma-1}\frac{zu'}{u}\right]\,.
\end{equation}
We may rewrite (\ref{dusg}) in the form
\begin{equation}
\label{Lsg}
{\frac{d}{dz}}\left[{\ts{1\over2}}u'^{2}+U(u,z)\right]
=D(u,u',z) \,.
\end{equation}
When $z\to\infty$, equation (\ref{dusg}) has a constant solution
$u_0=\left({3\over2}\gamma\right)^{(\gamma-1)/\gamma}$, which is 
the unique
minimum of $U(u,z)$. Thus, considering that $u$ lies in a 
neighbourhood of
$u_0$, we get from (\ref{Rsg}) that
\begin{equation} \label{Rasg}
R\approx 1-\frac{\gamma}{\gamma-1}\left(\frac{2}{3\gamma}\right)^
{1/\gamma}\left(\frac{k}{v}\right)^2zu' \,.
\end{equation}
We can see how $R\to 1$ when $z\to\infty$ through a linearization of 
(\ref{dusg}):
\begin{equation} \label{dulsg}
u''+2(q-1)\Omega  u'+\Omega 
\frac{\left(u-u_0\right)}{z}=0 \,,
\end{equation}
where $\Omega =3(\gamma/\a) v^2(3\gamma/2)^{q-2}(q-1)^{-2}$.
By 
the change of variable $u=u_0+w\exp[(1-q)\Omega  z]$, 
equation (\ref{dulsg}) becomes
\begin{equation} \label{dw}
w''+\left[-(q-1)^2\Omega ^2+\frac{\Omega }{z}\right]w=0 \,,
\end{equation}
which can be solved in terms of confluent hypergeometric 
functions. It has
the following leading behaviour for $z\to\infty$:
\begin{equation} \label{wz}
w(z)\sim C_1 z^{1/2(1-q)}\exp\left[(q-1)\Omega  z\right]
+C_2 z^{1/2(q-1)}\exp\left[(1-q)\Omega  z\right] \,,
\end{equation}
where $C_{1,2}$ are arbitrary integration constants.  
Thus $z u'\sim
z^{1/2(1-q)}\to 0$ for $z\to\infty$ and $q>1$, and we get the 
expansions
\begin{equation} \label{Uu}
U(u,z)=\frac{3\left(\gamma-1\right)^2 v^2}{\left(q-1\right)^2\alpha}
\left[\frac{u^{(\gamma q-2)/(\gamma-1)}}{\gamma q-2}-
\frac{3\gamma u^{(\gamma  q-2-\gamma)/(\gamma-1)}}
{2\left(\gamma q-2-\gamma\right)}\right]\frac{1}{z}+
O\left(\frac{1}{z^2}\right) \,,
\end{equation}
for $\gamma q\neq 2$ or $2+\gamma$,
\begin{equation} \label{Uu2}
U(u,z)=\frac{3\left(\gamma-1\right) v^2}{\left(q-1\right)^2\alpha}
\left[\ln \frac{u}{u_0}+\frac{3\left(\gamma-1\right)}{2}
u^{\gamma/(1-\gamma)}\right]\frac{1}{z}+
O\left(\frac{1}{z^2}\right) \,,
\end{equation}
for $\gamma q=2$,
\begin{equation} \label{Uu3}
U(u,z)=\frac{3\left(\gamma-1\right) v^2}{\left(q-1\right)^2\alpha}
\left[\frac{\gamma-1}{\gamma}u^{\gamma/(\gamma-1)}-
\frac{3\gamma}{2}\ln \frac{u}{u_0}\right]\frac{1}{z}+
O\left(\frac{1}{z^2}\right) \,,
\end{equation}
for $\gamma q=2+\gamma$, and
\begin{equation} \label{D}
D(u,u',z)=-\left[\frac{6\gamma v^2 }
{\alpha\left(q-1\right)}\right]u^{\gamma(q-2)/(\gamma-1)}u'^2+
O\left(\frac{1}{z}\right) \,.
\end{equation}

For mathematical completeness, we will consider
also the case $\gamma=1$. We make the change of variables
\[
H^{-1} = t\exp\left[-u(z)\right]\,,~ z=t^{q-1} \,,
\]
in (\ref{14}), which then takes the form
\begin{eqnarray}
&& u''+\frac{1}{2}\left\{\frac{3v^2}{R\alpha}e^{-\left(q-1\right)u}
+\left[\left(q-2\right)e^{-u}+3\left(\frac{k^2}{R}+1\right)\right]
\frac{1}{z}\right\}\frac{u'}{q-1} \nonumber\\
&&{}+
\frac{3v^2}{4\left(q-1\right)^2R\alpha}\left[3e^{-\left(q-2\right)u}-
2e^{-\left(q-1\right)u}\right]\frac{1}{z} \nonumber\\
&&{}+
\frac{1}{4\left(q-1\right)^2}\left\{2e^{-u}+
9\left[\frac{k^2-v^2}{R}+\frac{1}{2}\right]e^u-
12\left(1+\frac{k^2}{R}\right)\right\}\frac{1}{z^2}=0 \,.
\label{du1}
\end{eqnarray}
When $z\to\infty$, equation (\ref{du1}) has a constant solution
$u_0 =\ln{3\over2}$. Again, we may write (\ref{du1}) in the form
(\ref{Lsg}), and find that $u_0$ is the unique
minimum of $U(u,z)$. Thus we get
\begin{equation} \label{Uugamma1}
U(u,z)=\frac{3v^2}{4\alpha\left(q-1\right)^2}\left[\frac{2}{q-1}
e^{-\left(q-1\right)u}
-\frac{3}{q-2}e^{-\left(q-2\right)u}\right]\frac{1}{z}+
O\left(\frac{1}{z^2}\right)\,,
\end{equation}
for $q\neq 2$,
\begin{equation} \label{Uugamma12}
U(u,z)=\frac{3v^2}{4\alpha}\left(2e^{-u}-3u\right)\frac{1}{z}+
O\left(\frac{1}{z^2}\right)   \,,
\end{equation}
for $q=2$, and
\begin{equation} \label{Dgamma1}
D(u,u',z)=-\frac{3v^2}{2\alpha
\left(q-1\right)}e^{-\left(q-1\right)u}u'^2+
O\left(\frac{1}{z}\right) \,.
\end{equation}

As $U(u,z)$ has a unique minimum at $u_0$ for any $q\neq 1$,
and $D(u,u',z)$ is negative definite for $q>1$, we find that solutions
with leading Friedmann behaviour $a\sim t^{2/(3\gamma)}$ when 
$t\to\infty $ are
asymptotically stable for $q>1$.


\section{Models with ideal-gas temperature}

\subsection{The de Sitter solution}

Using the ideal gas equation $p=nT$ instead of the power-law
for the temperature one obtains (\ref{15}).
Setting $H=H_0$ in (\ref{15}) then gives
\[
H_0 ^{1-q}=\frac{\alpha}{v^2\gamma}(v^2-k^2) \,.
\]
The upper bound (\ref{13'}) on the bulk stress implies
the limit $k\leq v$. Thus de Sitter solutions may exist.
On the other hand the specific 
entropy may be calculated and one obtains (for $\gamma\neq1$)
\[
s(t)= s_0 + 3H_0\left( \frac{\gamma}{\gamma -1}\right)t \,,~
\se(t)=s(t)-{\gamma\over v^2(\gamma-1)} \,.
\]
Linear analysis around the fixed point 
$(0,H_0)$ gives the following
information: for $1-\omega_1 \leq q<1$, it is an asymptotically 
stable node; for $q>1$, it is a saddle point; for $q<1-\omega_1$, 
it is an asymptotically stable focus, where
\[
\omega _1 = \frac{v^6}{2\gamma (v^2-k^2)^2}>0 \,.
\]
Thus the solution is an attractor for $q<1$ (as in the 
barotropic temperature case \cite{mm}).
For $q=1$, de Sitter inflation occurs 
for $H_0$ arbitrary, provided that
$\alpha = v^2\gamma/(v^2-k^2)$.

\subsection{Solution for $q=1$}

When $q=1$, power-law solutions $a\sim t^N$ exist if,
by (\ref{15}),
\begin{equation}
N= \left(\frac{2}{3\gamma}\right) \frac{\alpha k^2+v^2\gamma}{\alpha
(k^2-v^2)+v^2\gamma}= \frac{2}{3\gamma}
\left[1+\frac{\alpha}{\gamma} + O(\alpha ^2)\right] \,.
\label{2p}
\end{equation}
For positive entropy generation we require
\begin{equation}
k\leq \left(\frac{3\gamma N}{3\gamma N-2}\right)^{1/2} v \,,
\label{3rc}
\end{equation}
but it is easy to show that this restriction is always fulfilled for
(\ref{2p}) with arbitrary $v$, $k$ and $\alpha$.
From (\ref{1}) and (\ref{2}) with $\gamma \neq 1$, we find
\[
s(t) = s_0+\left(\frac{3\gamma N-2}{\gamma -1}\right)\ln t \,,~
\se(t) = s(t)-{(3\gamma N-2)^2\over 18N^2v^2\gamma(\gamma-1)}\,,
\]
so that positive $s$ requires $N>2/3\gamma$ -- which 
is always fulfiled for expanding solutions,
as one can see from (\ref{2p}). These solutions ($N>0$) require
\[
k^2>\left({\a-\gamma \over \a}\right)v^2\,.
\]
In order to get an
inflationary power-law solution, it is necessary that 
$\alpha >\alpha _{\rm min}$ where 
\[
\alpha _{\rm min}= \frac{v^2\gamma(3\gamma -2)}{3\gamma(v^2-k^2)+2k^2}
\,.
\]

For $q=1$, and assuming that $k^2/v^2\equiv \epsilon^2 \ll 1$ and
$\dot H/H^2$ is bounded, we can isolate $\ddot{H}$ in (\ref{15}) 
to first order in $\epsilon^2$:
\begin{eqnarray} 
&& H\ddot H+\frac{3\gamma v^2}{\alpha}
\left(1+2\epsilon^2\right)H^2\dot H+
2\left(\frac{k^2}{\alpha}-1\right)\dot H^2 \nonumber\\
&&{}+\frac{9}{2}v^2\gamma
\left[-1+\frac{\gamma}{\alpha}\left(1+\epsilon^2\right)\right]H^4=0\,.
\label{Hig}
\end{eqnarray}
With the change of variables $H=y^n$, $t=\a\theta/
[3\gamma v^2(1+2\epsilon^2)]$,
where $n=\a/(2k^2-\alpha)$, equation (\ref{Hig}) takes exactly
the same form as
(\ref{y}), with
\begin{equation} \label{betaig}
\beta=\frac{\epsilon^2\left(\gamma-\alpha+\gamma\epsilon^2\right)}
{\gamma\left(1+2\epsilon^2\right)^2} \,.
\end{equation}
Then, making the same nonlocal transformation (\ref{ch}) as before,
equation (\ref{y}) is again brought to the linear form (\ref{z}).
From (\ref{betaig}) we find that $1-4\beta>0$, so we
obtain the general solution of (\ref{Hig}) in parametric form
\begin{eqnarray}
H(\eta)&=&\left\{(n+1)\left[C_+\exp\left(\lambda_+\eta\right)+
C_-\exp\left(\lambda_-\eta\right)\right]\right\}^{\a/2k^2} \,,
\label{gsig0}\\
t(\eta)&=&t_0+\left[\frac{\a}{3\gamma v^2\left(1+2\epsilon^2
\right)}\right]\int {d\eta\over H(\eta)}\,,
\label{gsig}
\end{eqnarray}
where $\lambda_\pm=[-1\pm\left(1-4\beta\right)^{1/2}]/2$ are real,
and $t_0$ and $C_\pm$ are arbitrary integration constants. 
When either of $C_\pm$
vanishes, we obtain one-parameter 
families of solutions, which have the
form:
\begin{eqnarray}
H_{\pm }(t)&=&{\nu _{\pm }\over(t-t_0)}~\mbox{ if }~
\alpha \neq \alpha_0
\equiv \gamma\left(1+{k^2\over v^2}\right)\,,
\label{1paig}\\
H_{-}&=&{\nu _0\over(t-t_0)}\,,~ H_{+}=H_0~\mbox{ if }
~ \alpha = \alpha_0\,,
\label{1pbig}
\end{eqnarray}
where
\begin{eqnarray}
\nu_{\pm}&=& \frac{1+2\epsilon^2}
{3\left(\gamma-\alpha+\gamma\epsilon^2\right)}
\left\{1\pm\left[{\gamma+4\epsilon^2\a \over \gamma\left(1+2
\epsilon^2\right)^2}\right]^{1/2}\right\} \,,
\label{1pcig}\\
\nu_0 &=& \left(\frac{2}{3\gamma}\right)
\frac{\epsilon^2}{1+2\epsilon^2}\,.
\label{1pdig}
\end{eqnarray}
The specific
entropy $s$ is given by (\ref{1se}) and
$\se$ by (\ref{2sef}), and the restrictions (\ref{1r}) and
(\ref{2r}) also apply here, with $\nu_{\pm}$ and $\nu _0$ given by
(\ref{1pcig}) and (\ref{1pdig}),
and satisfying similar restrictions to the barotropic temperature
case. Note that $\a_0$
is the value of $\alpha$ which
arises from the requirement that
 the solution of (\ref{Hig}) is de Sitter, and it
can be also obtained as the first order expansion of the expression
$\alpha_0=\gamma/\left(1-\epsilon^2\right)$ that arises
from (\ref{15}).

On the other hand,
provided the parameters satisfy the constraint
\begin{equation} \label{mag}
\phi=\frac{1-8\psi\epsilon^2+16\psi^2\epsilon^4}
{1+2\psi+\epsilon^2-4\psi^2\epsilon^2}\,,
\end{equation}
where $\phi\equiv \gamma/\alpha$ and $\psi\equiv v^2/\alpha$,
the general solution of (\ref{Hig}) takes
 an explicit form, and we get for the
scale factor
\begin{equation} \label{atig}
a(t)=a_0\left|\left|t-t_0\right|^{n+1}+K\right|^{\mu}\,,
\end{equation}
where
$K$ is arbitrary and
\begin{equation} \label{mu}
\mu=\frac{4k^2-\alpha}{3\gamma v^2\left(1+2\epsilon^2\right)}\,.
\end{equation}
We have checked that the constraint
 (\ref{mag}) is satisfied for suitable
values of the parameters. For instance,
if $\epsilon<0.1$, then $\phi<1$.
This constraint also reduces, in the limit $\epsilon\to 0$,
 to the constraint
on the parameters found in a previous paper for ideal-gas 
temperature in
the linear theory \cite{cjm}.
It must be remarked however, that the form of the solutions
is quite different. This shows that, in general, the solutions are not
analytic in $k$ for $k\to 0$. In other words, this limit is singular.

The requirement that $\dot H/H^2$ remains bounded means that not
all the
two-parameter families of solutions (\ref{gsig})
of equation (\ref{Hig})
are also (approximate) solutions of (\ref{15}). 
Those that satisfy it behave as follows:
The evolution begins at a singularity, with Friedmann leading
behaviour as in (\ref{1paig}--) for $\alpha\neq \alpha_0$,
 or (\ref{1pbig}--) for
$\alpha=\alpha_0$. Then the expansion becomes

1. asymptotically Friedmann, as in (\ref{1paig}+),
 for $\alpha<\alpha_0$;

2. asymptotically de Sitter, as in (\ref{1pbig}+),
 for $\alpha=\alpha_0$;

3. divergent at finite time, with leading behaviour (\ref{1paig}+) for
$\alpha>\alpha_0$.
The relationships for $s$ and $\se$ and the the thermodynamical
constraints (\ref{1r}) and (\ref{2r})
in the limit $ t\rightarrow \infty$,
are the same, but with $\nu_\pm$ replaced by $\mu (n+1)$.

\subsection{Asymptotic analysis for $q>1$ and stability}

For $q>1$ it is easy to check that (\ref{15}) admits a solution whose leading
term is $2/(3\gamma t)$.  To study its stability, we make  the change of
variables
\[
H^{-1} = tu(z)\,,~~ t^{q-1} = z \,,
\]
in (\ref{15}), which then becomes
\begin{eqnarray}
&& u''+\left(\frac{6\gamma v^2}{\alpha R}u^{q-2}
+\frac{q}{z}\right)\frac{u'}{q-1} \nonumber\\
&&{}+\frac{3\gamma v^2}{\left(q-1\right)^2\alpha Rz}
\left(u^{q-1}-\frac{3\gamma}{2}u^{q-2}\right)+
\frac{9\gamma v^2}{2\left(q-1\right)^2 z^2 u}=0 \,,
 \label{dus}
\end{eqnarray}
where
\[
R=1-\epsilon^2+\frac{2\epsilon^2}{3\gamma }
\left[u+\left(q-1\right)zu'\right] \,.
\]
We rewrite (\ref{dus}) in the same form (\ref{Lsg}) as before.
When $z\to\infty$, equation (\ref{dus}) has a constant solution
$u_0={3\over2}\gamma$, which is the unique
minimum of $U(u,z)$. Thus, considering
that $u$ lies in a neighbourhood of
$u_0$, we get
\[
R\approx 1+\frac{2}{3\gamma}\epsilon^2\left(q-1\right)zu'\,.
\]
We can see how $R\to 1$ when $z\to\infty$ through a linearization
of (\ref{dus}), which again leads to (\ref{dulsg}), with
 $\Omega=3(\gamma/\a)
v^2(q-1)^{-2}(3\gamma/2)^{q-2}$. By 
the same change of variable $u=u_0+w\exp[(1-q)\Omega z]$,
equation (\ref{dulsg}) is again brought to the form
(\ref{dw}). As before, this
can be solved in terms of confluent hypergeometric
functions, and has
the same leading behaviour (\ref{wz}) for $z\to\infty$.
Thus $z u'\sim
z^{1/2(1-q)}\to 0$ for $z\to\infty$ and $q>1$,
and we get the expansions
\begin{eqnarray*}
D(u,u',z)&=&-\frac{6\gamma v^2 u^{q-2}}
{\alpha\left(q-1\right)R}u'^2+
O\left(\frac{1}{z}\right)\,,               \\
U(u,z)&=&\frac{3\gamma v^2}{\left(q-1\right)^2\alpha}
\left[\frac{u^q}{q}-\frac{3\gamma}{2\left(q-1\right)}u^{q-1}\right]
\frac{1}{z}+O\left(\frac{1}{z^2}\right)\,.
\end{eqnarray*}
As $U(u,z)$ has a unique minimum at $u_0$ for any $q\neq 1$,
and $D(u,u',z)$ is negative definite for $q>1$, we
conclude that solutions
with leading Friedmann behaviour $a\sim t^{2/(3\gamma)}$
when $t\to\infty$ are
asymptotically stable for $q>1$.

\newpage

\section{Concluding remarks}

With the aim of treating
 dissipative processes that do not remain close
to equilibrium, we have used the nonlinear generalization of the
standard causal theories proposed in \cite{mm}. Barotropic and 
ideal-gas forms for the temperature have been used to close the
dynamical equations of the universe.

The evolution of the universe is qualitatively similar in
both cases, and we can summarize the
main results as follows. Singular
solutions are found for $q=1$ with a power-law and inflationary
behaviour in the limit $t\to\infty$.
The initial and final
behaviour of the solutions follows closely
that of the linear theory with barotropic temperature, in
the `full' causal as well as the `truncated' causal
 and the noncausal cases.
The logarithmic behaviour found for ideal-gas temperature
in \cite{cjm} no longer
appears when  the nonlinear transport equation is considered.
Thermodynamic restrictions,
arising from positivity of the entropy, have been also derived.
The evolution begins at a singularity with a Friedmann leading
behaviour, and the future dynamic is asymptotically Friedmannian,
asymptotically de Sitter or divergent at finite time, for specific
values of the bulk viscosity constant $\alpha$.

The behaviour of particular de Sitter
solutions has already been analyzed in
\cite{mm}, for barotropic temperature,
 and it is also derived here for the ideal-gas temperature.
 For $q<1$, a stable exponential inflationary phase occurs in the
far future when $ 1\le\gamma<2$.
The condition $\alpha=\alpha_0$ is required
if $q=1$, and such behaviour is unstable for $q>1$.
Our results show that occurrence of viscosity-driven
exponential inflationary
behaviour depends mainly on the equation of state, rather than on the
thermodynamical theory employed.

For $q>1$ the perfect fluid behaviour $a\sim t^{2/(3\gamma)}$ when
$t\rightarrow\infty $ is asymptotically stable, because the viscous
pressure
decays faster than the thermodynamical pressure.
 However, if $q=1$ and
$\alpha<\alpha_0$, both pressures decay asymptotically
as $t^{-2}$, and the
exponent becomes $\nu_+$. The perfect fluid behaviour
becomes unstable if
$q<1$.

In essence we have analyzed the existence of specific behaviour for a
dissipative universe not close to equilibrium in terms of the
thermodynamic parameters related
to $\tau$, $q$ and the equations of state.

A final comment concerns the robustness of our results relative to the
spatial geometry and the source of the gravitational field. Firstly,
the question arises as to what the effect of non-zero spatial
curvature in the FLRW metric will be. In the case of non-causal
bulk viscosity, where $\Pi$ is {\em algebraically} determined
by $H$, it is relatively straightforward to investigate this
question \cite{b1,b2,b3,t}. By contrast, in the causal theory
(linear as well as nonlinear), $\Pi$ is not algebraically 
determined by $H$ but satisfies an evolution equation that couples it
differentially to the expansion. Curvature introduces the scale
factor $a=\exp\int Hdt$ explicitly into the Friedmann equation,
and makes it more difficult to decouple the equations. Thus the
effect of curvature in the causal case will be far more difficult
to determine in general. (In \cite{chm}, curvature is considered
in the linear causal theory, but with highly simplified equations
of state.)
This is a subject for further 
investigation.

Secondly, it could be interesting, but in general very difficult, to
determine the effect of introducing another fluid, without
bulk viscous stress. In the non-causal theory, this question
is much less difficult to investigate \cite{b2,b3}. In the causal 
theory, this is another subject for further investigation.

\ack

VM thanks the Spanish
Ministry of Education for partial support
under grant PB94-0718. LPC and ASJ thank the
University of Buenos Aires for partial support under project EX-260.


\end{document}